\documentclass{XrU2005}
\usepackage{natbib,graphics,graphicx}
\title{An X-ray source population study of the Andromeda galaxy M 31}
\author{W. Pietsch}
\affil{Max-Planck-Institut f\"ur extraterrestrische Physik, Giessenbachstra\ss e,
85741 Garching, Germany}

\def\cm-2{cm$^{-2}$}

\def\ein{{\sl Einstein}}
\def\ro{{\sl ROSAT}}
\def\xmm{{\sl XMM-Newton}}
\def\chandra{{\sl Chandra}}
\def\n253{NGC~253}
\def\m31{M~31}
\def\me33{M~33}
\def\mx7{M~33~X$-$7}
\def\x7{\hbox{X$-$7}}

\newcommand{\ergs}[1]{$\times10^{#1}$ \hbox{erg s$^{-1}$}}
\newcommand{\oergs}[1]{$10^{#1}$ erg s$^{-1}$}
\newcommand{\hcm}[1]{$\times10^{#1}$ cm$^{-2}$}

\newcommand{\expo}[1]{$\times10^{#1}$}

\newcommand{\nh}{\hbox{$N_{\rm H}$}}

\begin{document}

\keywords{galaxies: individual (M~31), novae, cataclysmic variables,
supernova remnants, X-rays: galaxies, X-rays: binaries,
X-rays: bursts}

\maketitle

\begin{abstract}
Archival 
\xmm\ EPIC observations reveal the population of X-ray sources of the 
bright Local Group spiral galaxy \m31, a low-star-formation-rate galaxy 
like the Milky Way,   
down to a 0.2-4.5 keV luminosity of 4.4\ergs{34}. With the help of X-ray 
hardness ratios and
optical and radio information different source
classes can be distinguished. The survey detected 856 sources in an area of 1.24  square degrees.
Sources within \m31\ are 44 supernova remnants (SNR) and candidates, 18
super-soft sources (SSS), 16 X-ray binaries (XRBs) and candidates, as well as 37
globular cluster sources (GlC) and candidates, i.e. most likely low mass  XRBs
within the GlC. 567 hard sources may either be  XRBs or Crab-like SNRs in \m31\
or background AGN. 22 sources are new SNR candidates in \m31\ based on X-ray
selection criteria. Time variability information can be used to improve the
source classification. Two GlC sources show type I X-ray bursts as known from
Galactic neutron star low mass XRBs. Many of the \m31\ SSS detected with \xmm, \chandra\
and \ro, could be identified with optical novae. Soft X-ray light curves can
be determined in \m31\ center observations for several novae at a time opening a new area of nova research.
\end{abstract}

\section{Introduction}
In the \xmm\ survey of the Local Group Sc galaxy \me33\
\citep[][hereafter PMH2004]{2004A&A...426...11P}, 
408 sources were detected in a 0.8 square degree field combining the counts 
of all
EPIC instruments, which could be identified and classified using  X-ray 
colors and time variability as well as optical and radio information. This proved to be an 
efficient way to separate super-soft X-ray sources (SSSs) and  thermal supernova 
remnants (SNRs) in  \me33\ from Galactic stars in the foreground and ``hard" 
sources. These hard sources may be either X-ray binaries (XRBs) or Crab-like 
SNRs in \me33\ or active galactic nuclei (AGN) in the background of the galaxy. 
The success of this survey inspired us for a similar analysis of all archival
\xmm\ observations of \m31. 

The Andromeda galaxy \m31\ is located at a distance similar to the one of
\me33\ \citep[780 kpc,][ i.e.
1\arcsec\ corresponds to 3.8 pc and the flux to  luminosity conversion factor
is 7.3\expo{49} cm$^2$]{1998AJ....115.1916H,1998ApJ...503L.131S} and -- compared
to the near face-on view of \me33\ -- is seen under
a higher inclination (78$^{\circ}$). 
The optical extent of the massive SA(s)b galaxy
can be approximated by an inclination-corrected $D_{25}$
ellipse with a large diameter  of 153.3' and axis ratio of 3.09
\citep{1991trcb.book.....D,1988ngc..book.....T}. With its moderate Galactic
foreground absorption  \citep[\nh = 7\hcm{20},
][]{1992ApJS...79...77S}, \m31\ is well suited to study the X-ray source
population and diffuse emission in a nearby spiral similar to the Milky Way.
\m31\ was a target for many previous imaging X-ray missions. The \ein\ 
Observatory detected 108 individual X-ray sources brighter than 5\ergs{36} 
\citep[see e.g.][]{1991ApJ...382...82T}. The sources were identified with young stellar associations, globular clusters 
(GCs) and SNRs. The \ro\ HRI \citep{1993ApJ...410..615P} detected
86 sources brighter than $\sim$\oergs{36} in the central area of \m31.
The \ro\ PSPC covered 
the entire galaxy twice in surveys conducted one year apart and detected altogether 
560 X-ray sources down to a limit of $\sim$5\ergs{35} and SSS were established as a new class 
of \m31\ X-ray sources 
\citep[][]{1997A&A...317..328S,2001A&A...373...63S}. 
The flux of many of the sources varied significantly between the 
\ein\ and \ro\ observations. Deep \chandra\ ACIS I and HRC observations of the central 
region (covered areas of 0.08 and 0.27 square degree) resolved 204 and 142 X-ray sources, respectively 
\citep{2000ApJ...537L..23G,2002ApJ...577..738K,2002ApJ...578..114K}.
A synoptic study of \m31\ with the \chandra\ HRC covered ``most" of the disk 
(0.9 square degree) in 17 epochs using short observations, and resulted in mean 
fluxes and long-term light curves for the 166 objects detected 
\citep{2004ApJ...609..735W}.
In these observations, several \m31\ SNRs were spatially resolved 
\citep{2002ApJ...580L.125K,2003ApJ...590L..21K} and bright XRBs in globular 
clusters and SSSs and quasi-soft sources (QSSs) could be characterized 
\citep[][]{2002ApJ...570..618D,2004ApJ...610..247D,2004ApJ...610..261G}.

One of \xmm's most important contributions to galaxy science was a deep
survey of the central region around the long axis of \m31\ as part of the 
guaranteed time program. This survey is unique
in that it has the greatest depth ($\sim$\oergs{35}) and best spatial resolution of any 
existing large area \m31\ survey. It has covered almost $3^\circ$ ($>$40 kpc) along the 
major axis of the galaxy and 30' ($\sim$7~kpc) along the central portion of the
minor axis. These deep \xmm\ observations have allowed
us for the first time to study the short-term time variability ($\sim$100~s and
shorter, see below) and spectra of bright X-ray sources 
in a galaxy outside the Milky Way and the Magellanic Clouds  
\citep[e.g.][]{2001A&A...378..800O,2003A&A...411..553B,2005astro.ph..8284B,
2004A&A...419.1045M}.
The observations revealed diffuse emission from the hot ISM in the centre
and the northern disk
\citep[][]{2001A&A...365L.195S,2001ApJ...563L.119T,TKP2004}, and were used 
to derive source luminosity distributions
\citep{2002ApJ...571L..17T}.

Here, I summarize results of our group mainly based on the archival \m31\ 
\xmm\ observations, including X-ray images and a source catalogue for the
archival  observations of \m31\ \citep[][hereafter PFH2005]{2005A&A...434..483P}, the detection 
of type I X-ray bursts in \m31\ \citep[][hereafter PH2005]{2005A&A...430L..45P} and on the detection
of optical novae in \m31\ as SSSs \citep[][hereafter PFF2005]{2005A&A...442..879P}.

\section{XMM-Newton survey of M 31}
\begin{figure}
\includegraphics[height=22.cm,clip]{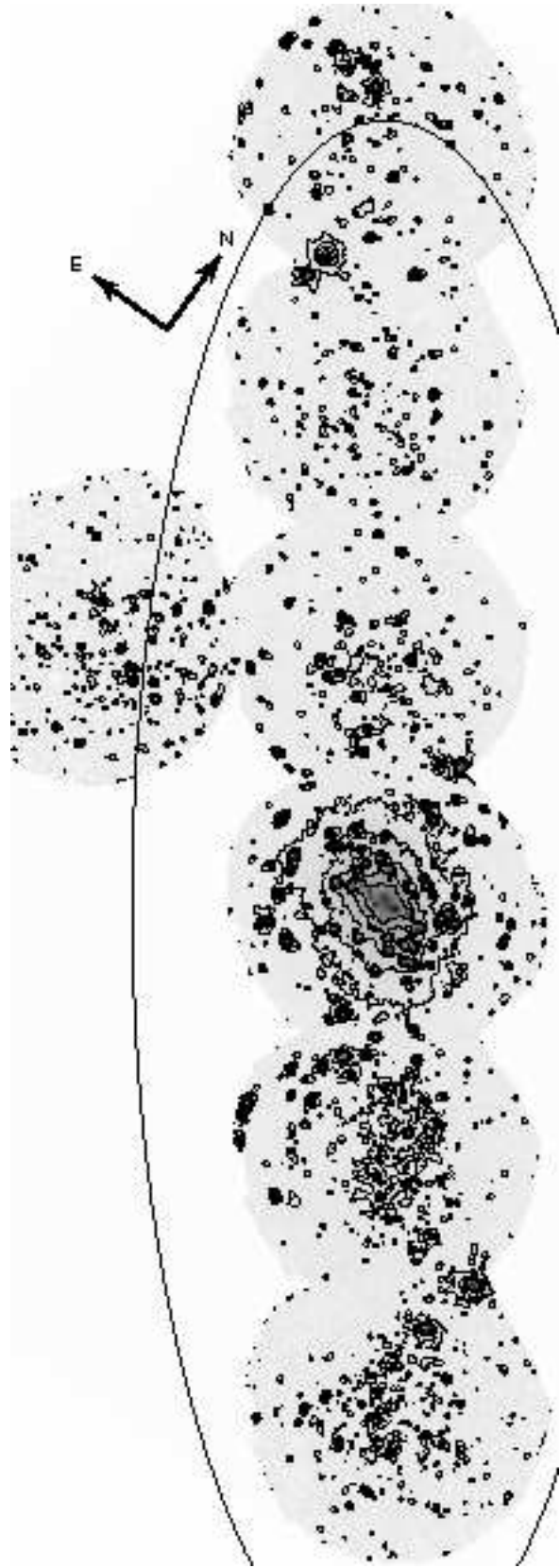}
  \caption{Combined \xmm\ EPIC image in the 0.2--4.5 keV band smoothed with a
  20" FWHM Gaussian. Orientation and optical $D_{25}$ ellipse are indicated.
     }\label{fig:m31}
\end{figure}

\begin{figure*}
\includegraphics[height=8.2cm,bb=50 90 355 460,angle=-90,clip]{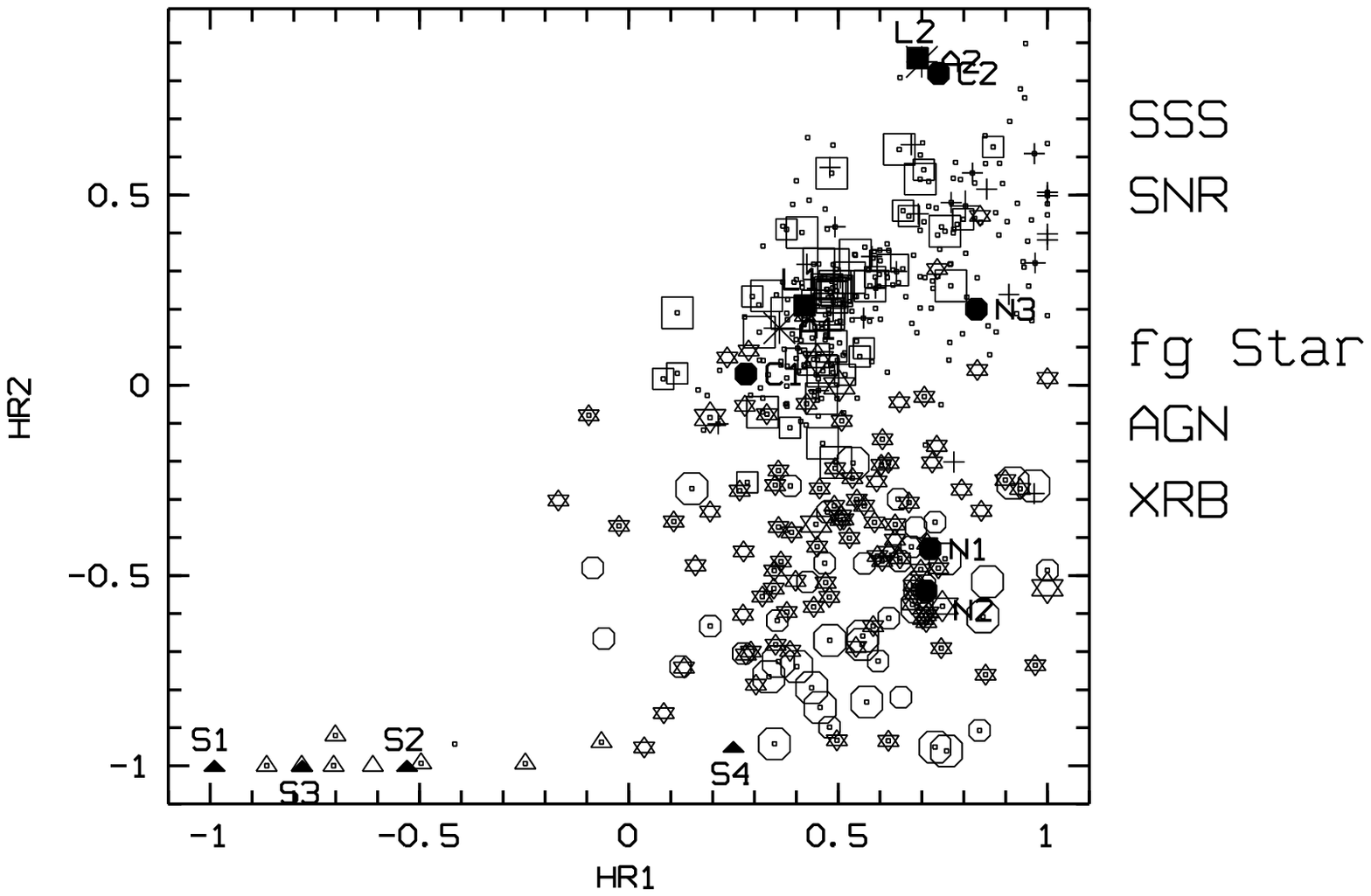}
\includegraphics[height=8.2cm,bb=50 90 355 460,angle=-90,clip]{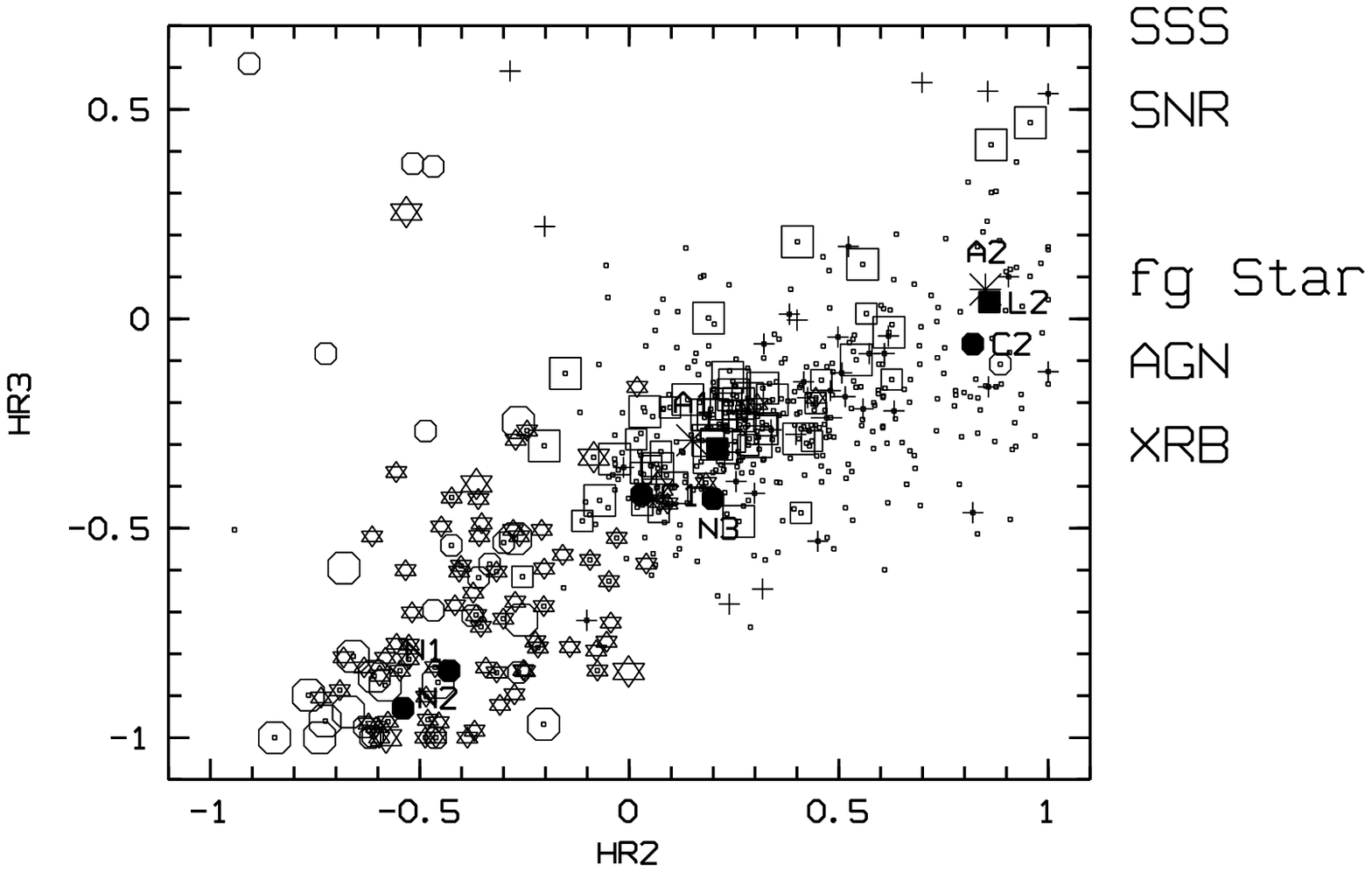}
  \caption{Hardness ratios (HR) detected by XMM-Newton EPIC.
      Shown as dots are only sources with HR errors smaller than 0.2 on both
      $HR{i}$ and  $HR{i+1}$. Foreground
     stars and candidate are marked as big and small stars, AGN and candidates as
     big and small crosses, SSS candidates as triangles, SNR and candidates as big and small
     hexagons, GlCs and XRBs and candidates as big and small squares. In addition, we mark positions derived from
     measured \xmm\ EPIC spectra and models  for SSSs (S1 to S4) as filled
     triangles, low mass XRBs (L1 and L2) as filled squares, SNRs (N132D as N1,
     1E~0102.2--7219 as N2, N157B as N3, Crab spectra as C1 and C2) as filled hexagons, 
     AGN (A1 and A2) as asterisk (extracted from Fig. 5 of PFH2005).
     }\label{fig:xhr}
\end{figure*}

PFH2005 have created merged medium and thin filter images for
the three EPIC instruments, in five energy bands (0.2--0.5 keV, 0.5--1.0 keV, 
1.0--2.0 keV, 2.0--4.5 keV, and 4.5--12 keV), using only times of low 
background from the archival \xmm\ \m31\ observations which at the time
contained four observations of the centre area of \m31\ separated by
half a year, two pointings in the southern, three in the northern disk, and 
one short observation in the halo, which all at least partly cover the optical
$D_{25}$ ellipse. In total, the observations in the analysis cover an area of 
1.24 square degrees (see Fig.~\ref{fig:m31}) with a limiting sensitivity of 
4.4\ergs{34} in the 0.2--4.5~keV band which 
is a significant improvement
compared to the \chandra\ surveys. However, up to now only about 2/3 
of the optical \m31\ extent ($D_{25}$ ellipse) are covered with a rather inhomogeneous 
exposure. There were still significant offsets 
between the observations that had to be corrected for before
merging. For the centre observations these offsets were
determined from source lists of the individual observations using the
USNO-B1, 2MASS, and \chandra\ catalogues to define an absolute reference frame.
This finally resulted in a residual systematic position error of less than 
0.5". The source detection procedures revealed 856 sources using simultaneously $5\times 3$
images (5 energy bands and PN, MOS1 and MOS2 camera). For the pointings into 
the disk and halo of \m31\ this procedure was applied to the individual observations. 
The centre pointings strongly overlap and therefore the images were merged to reach 
higher detection sensitivity.

Hardness ratios were calculated only for sources for which at 
least one of the two band count rates had a significance greater than $2\sigma$
(Fig.~\ref{fig:xhr}).  
In search for identifications, the X-ray source positions were correlated
with sources in the SIMBAD and NED archives
and within several catalogues. The cataloged X-ray sources are ``identified" 
or ``classified" based on properties in X-rays 
(hardness ratios (HR), variability, extent) and of correlated objects in other 
wavelength regimes. A source is counted as identified,
if at least two criteria secure the identification. Otherwise, it is only counted as 
classified.

\begin{table}
\begin{center}
\caption[]{Summary of identifications and classifications of \xmm\ X-ray sources in
the \m31\ and \me33\ fields (see PFH2005 and PMH2004).}\vspace{1em}
\renewcommand{\arraystretch}{1.2}
\begin{tabular}{lrrrr}
\hline
&\multicolumn{2}{c}{M 31}&\multicolumn{2}{c}{M 33}\\
\multicolumn{1}{c}{Source type} & 
\multicolumn{1}{c}{ident.} &
\multicolumn{1}{c}{class.} & 
\multicolumn{1}{c}{ident.} &
\multicolumn{1}{c}{class.} \\ 
\hline
fg Star & 6 & 90 &5 &30\\
AGN  &   1 & 36 & &12\\
Gal  &   1 & & 1& 1 \\
GalCl  &  1 &1\\
SSS  &     &  18 & & 5\\
SNR  &   21 & 23  &21+2 &23-2\\
GlC  &   27 & 10 \\
XRB  &   7 & 9 &2\\
hard &    & 567 &&267\\
\hline\\
\end{tabular}
\label{class}
\end{center}
\end{table}

Table ~\ref{class} summarizes identifications and classifications according to
the \xmm\ catalogues of \m31\ and \me33. For the SNRs in \me33\ two  
new optical counterparts for soft X-ray SNR candidates from the \xmm\
list are indicated \citep[see][]{2005AJ....130..539G}.
Detection of strong time variability in follow-up analysis will certainly move many
objects from the ``hard" to ``XRB" classification.
Comparison to earlier X-ray surveys revealed transients not 
detected with \xmm, which add to the number of \m31\ XRBs. Up to to now, only
low mass X-ray binaries have been identified in \m31, mostly by correlations
with globular cluster sources. In \me33, however, besides the ultra-luminous
X-ray source close to
the nucleus (X--8, most likely a black hole XRB) the only other XRB identified
is the eclipsing high mass XRB X--7 with an orbital period of 3.45 d that has
been confirmed by the ellipsoidal heating light curve of its optical companion
\citep{2004A&A...413..879P}. 

\begin{figure}
\centering
\includegraphics[height=7.0cm,clip]{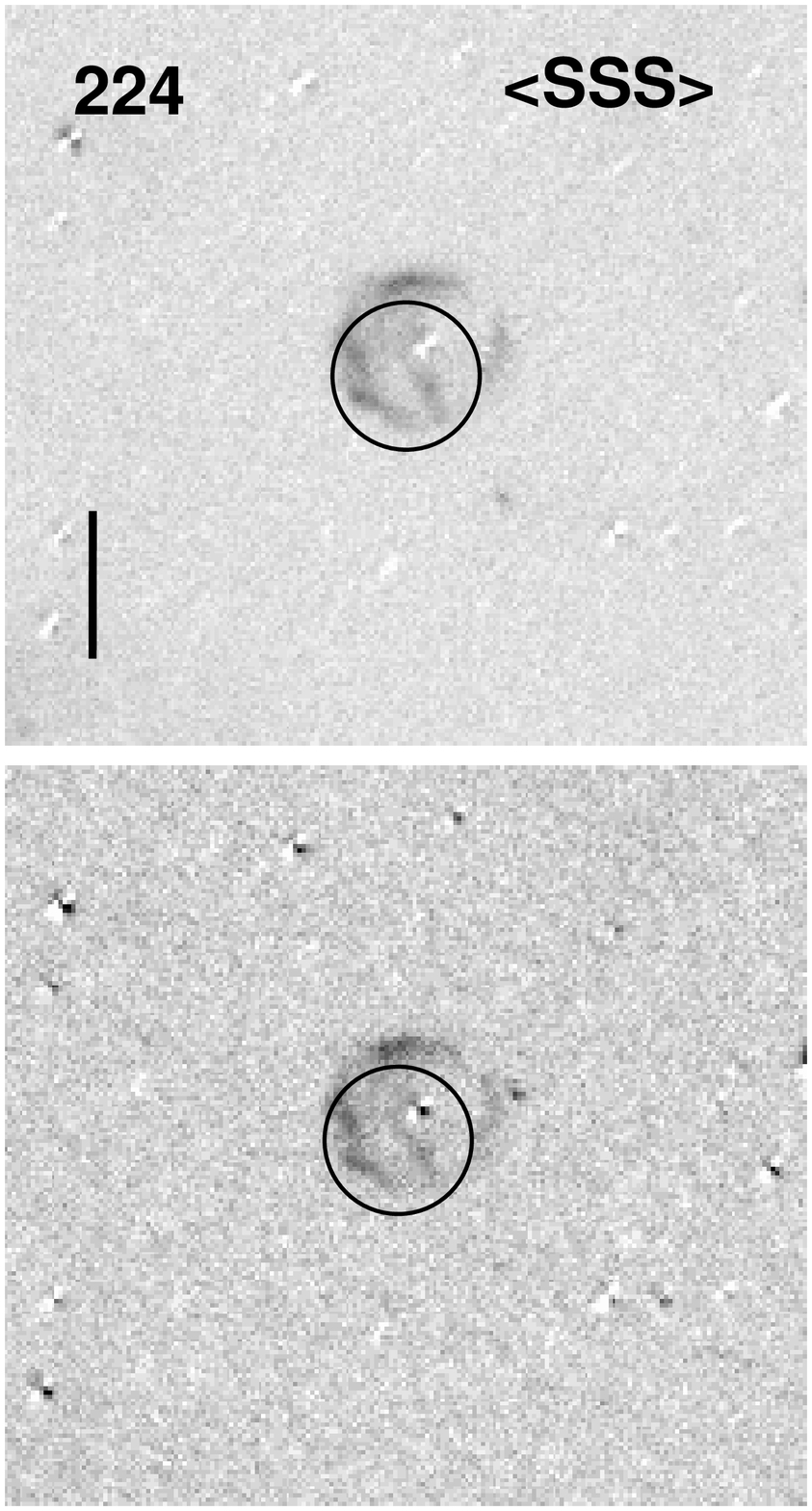}
\includegraphics[height=7.0cm,clip]{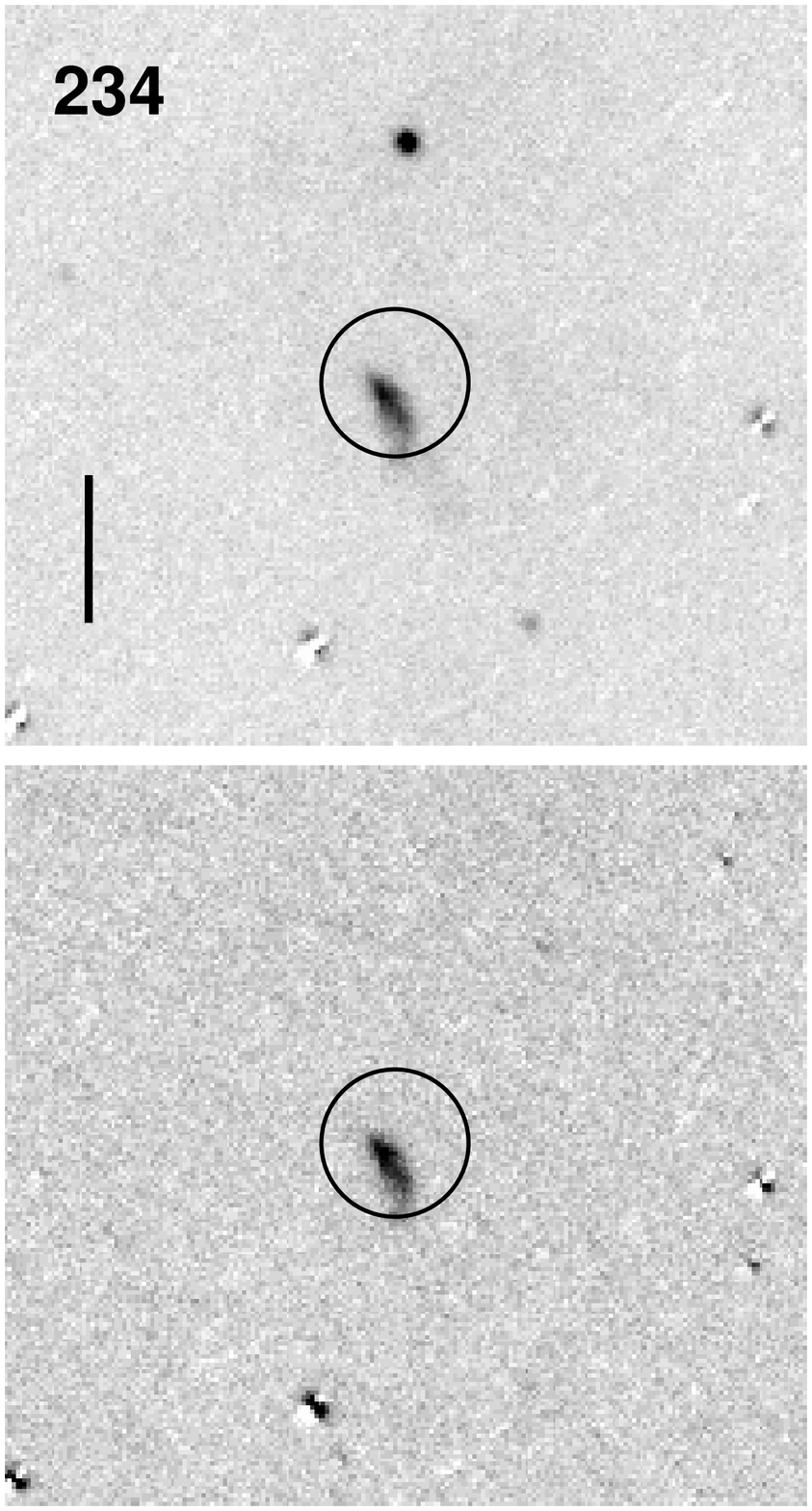}
\caption{SNR candidates from overlay of Local Group survey images (H$\alpha$
above, [SII] below): [PFH2005]~224 was classified as SSS candidate, 
[PFH2005]~234 was unclassified.
  \label{fig:snr}}
\end{figure}

Many foreground stars,
SSSs and SNRs can be classified or identified. The number of 44 SNRs and 
candidates more than 
doubles the X-ray-detected SNRs. 22 sources are new SNR candidates in \m31\
based on X-ray selection criteria.
Another SNR candidate may be the first plerion detected outside the Galaxy and the
Magellanic Clouds. Additional SNR candidates can be identified by comparing
cataloged X-ray sources with optical narrow filter images of the Local Group
survey of \citet{2001AAS...19913005M}. Figure~\ref{fig:snr} gives two examples
for two new optical SNR candidates proposed by X-ray source position and a
[S\,II]/H$\alpha$ ratio of the optical emission characteristic for optical SNRs.

\begin{figure}
\centering
\includegraphics[height=7.0cm,clip]{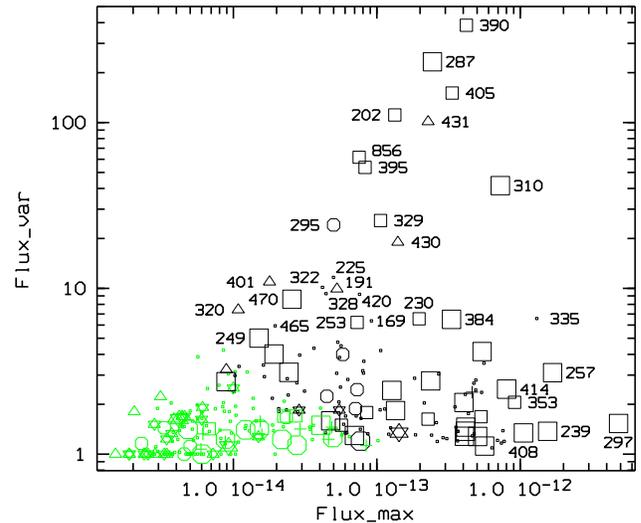}
\caption{Variability of X-ray sources within four overlapping \xmm\ 
observations to the \m31\ centre performed from June 2000 to January 2002.
  \label{fig:fluxvar}}
\end{figure}

\begin{figure*}
\centering
\includegraphics[height=8.2cm,clip]{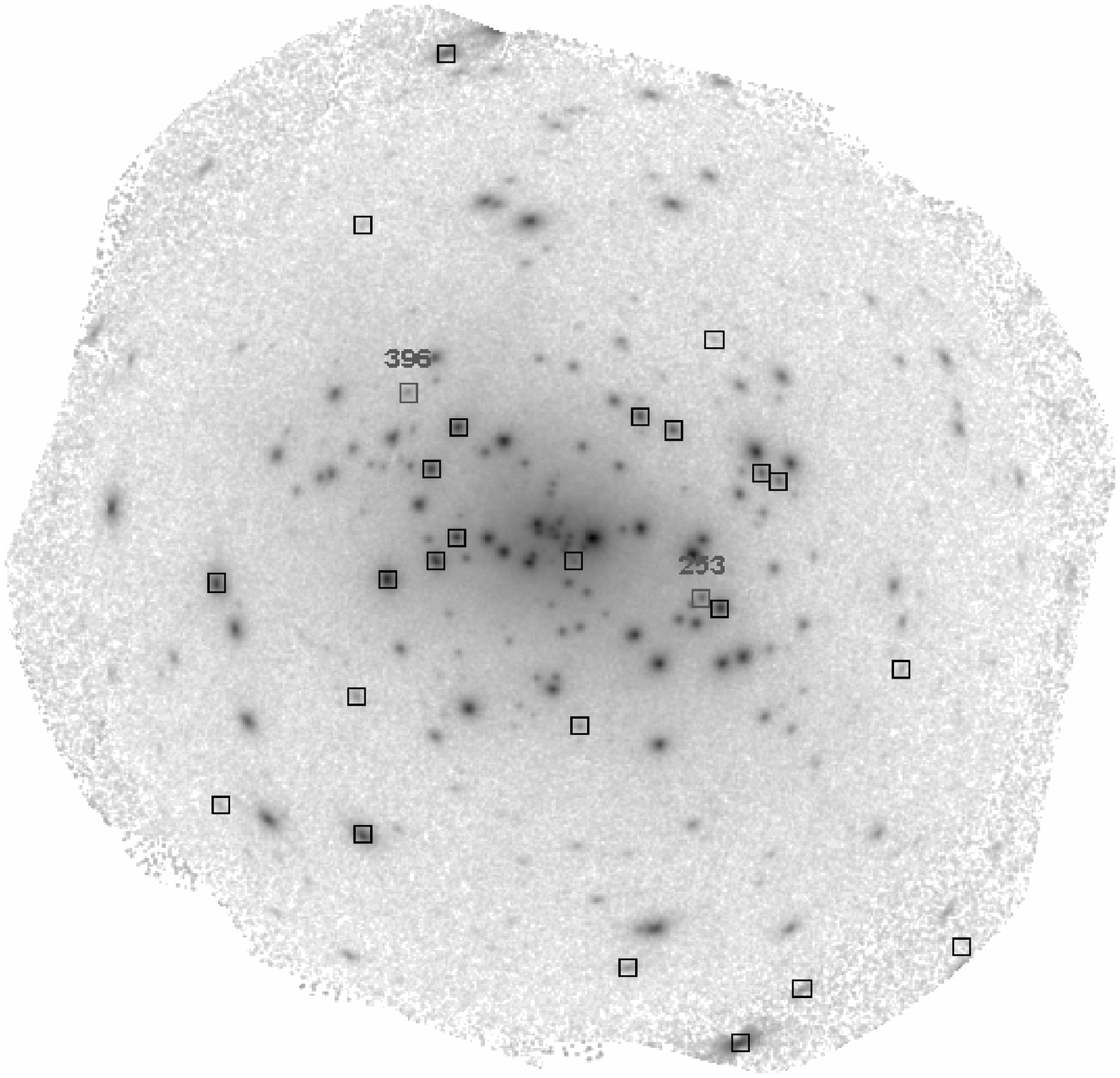}
\includegraphics[height=8.2cm,clip]{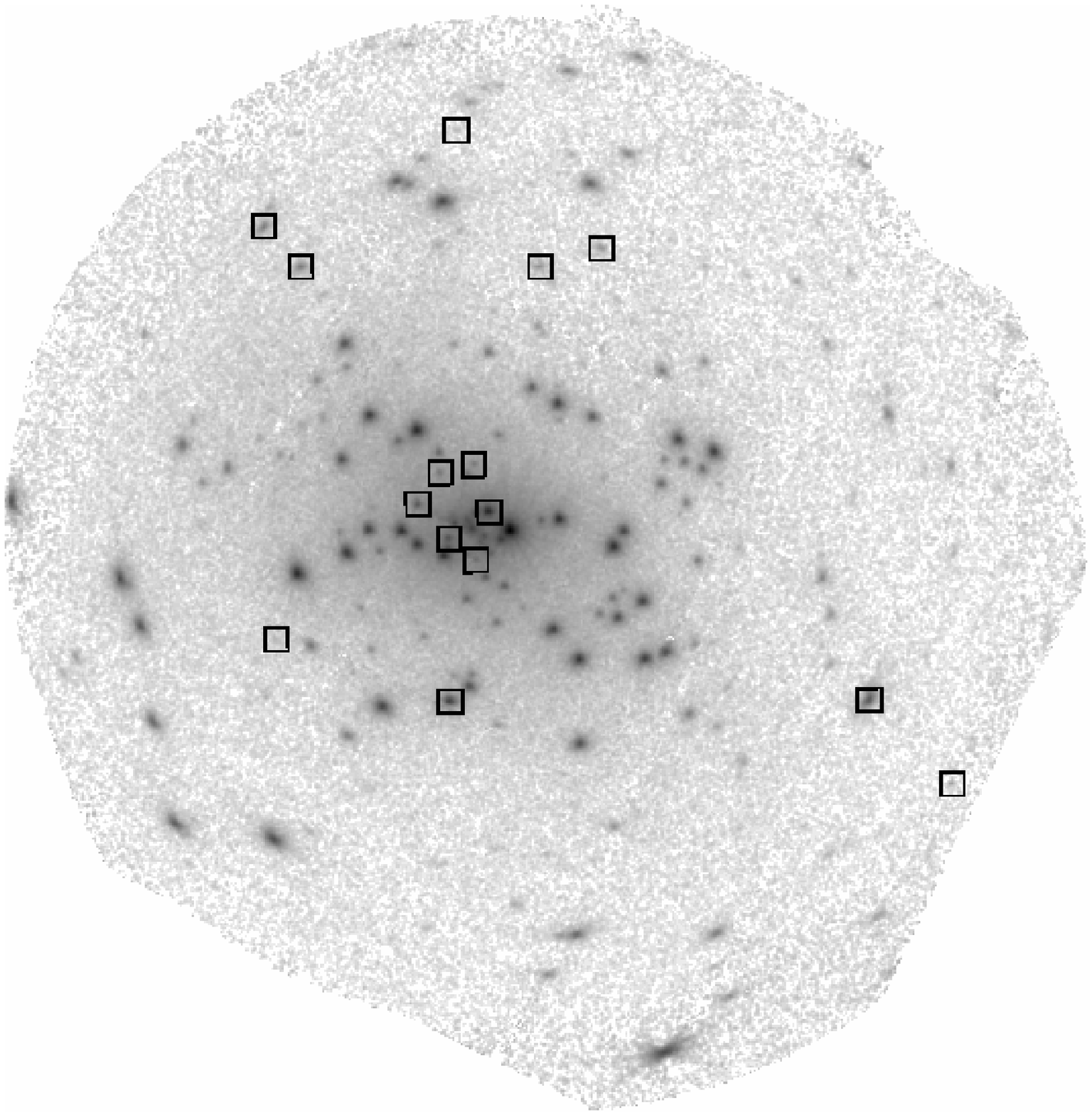}
\caption{Variability of X-ray sources between combined \xmm\ \m31 center
observations of June 2000 to January 2002 (left, globular cluster sources and
candidates marked, burst sources numbered) and observations 2.5 yr later
(right). New transients are marked. However, many bright sources from left 
image are missing.
  \label{fig:center}}
\end{figure*}

However, besides a few clearly
identified XRBs and AGN, and SNR candidates from positions in other
wavelengths, we have no clear hardness ratio criteria to select
XRBs, Crab-like SNR or AGN. They are all ``hard" sources (567 sources 
classified as hard in total). Only additional criteria like short or long term
X-ray variability or detailed spectral modeling will reveal their nature. Such
methods can be used in the \m31\ center area with four overlapping observations in
the \xmm\ archive (separated by half a year) and additional observations of the area
taken 2.5 yr later (see Fig.~\ref{fig:fluxvar} and \ref{fig:center}).

\section{Detection of type I X-ray burst sources in M 31}

\begin{figure}
\centering
\includegraphics[height=8.0cm,angle=-90,clip]{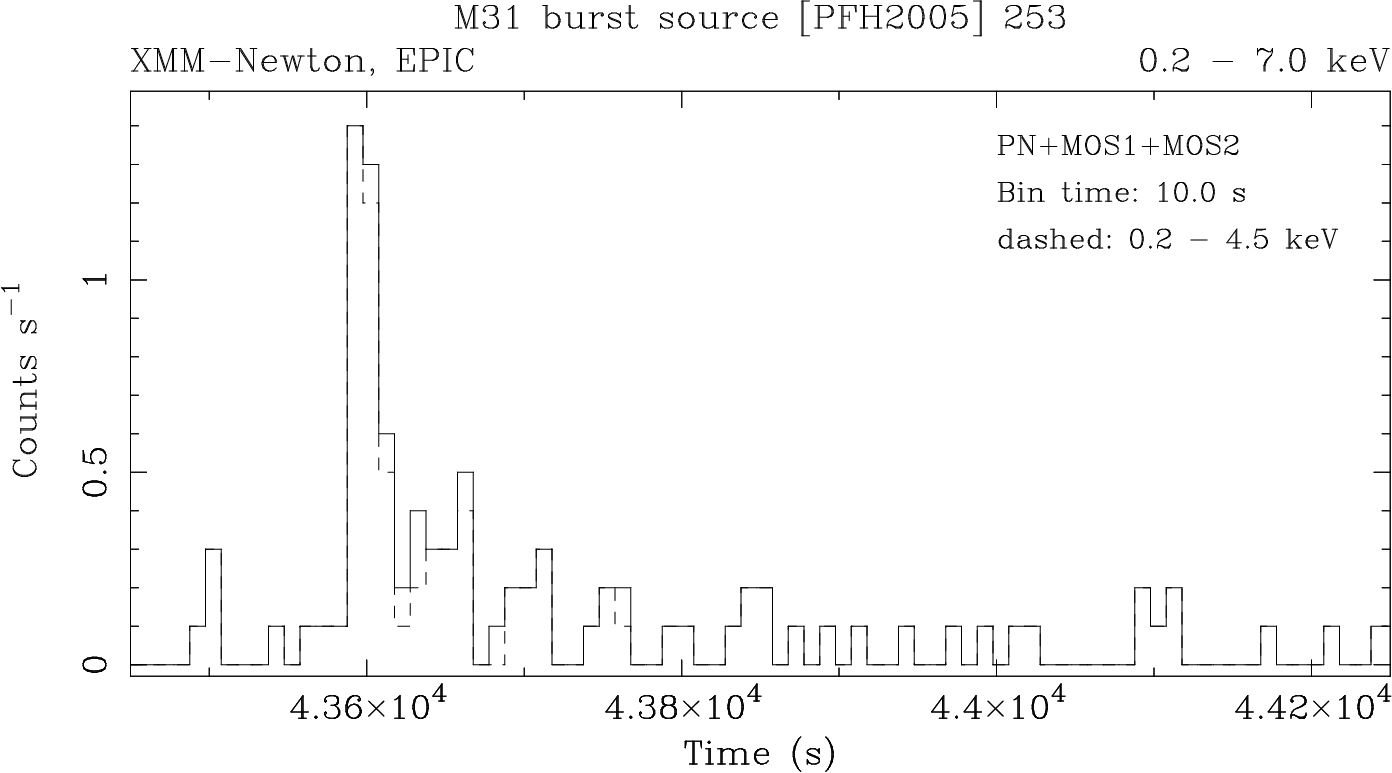}
\caption{Combined XMM-Newton EPIC light curve of a type I X-ray burst
  of source [PFH2005] 253 in M31 \citep[Fig.~3 from][]{2005A&A...430L..45P}.
  \label{fig:burst}}
\end{figure}

Within the Milky Way, bright globular cluster X-ray sources were identified as
low mass XRBs. Many of them show type I X-ray bursts identifying them as neutron
star systems. PH2005 searched for X-ray bursts in \xmm\ archival data of \m31\
sources which were identified or classified as globular cluster sources in the 
PFH2005 catalogue (Fig.~\ref{fig:center}).
Two bursts were detected simultaneously in EPIC pn and MOS detectors and some
more candidates in EPIC pn. The energy distribution of the burst photons and the
intrinsic luminosity during the peak of the bursts indicate that at least the
strongest events were type I radius expansion burst radiating during maximum 
at the Eddington limit of a 1.4 $M_{\odot}$ neutron 
star for hydrogen-poor matter (Fig.~\ref{fig:burst}). 
Standard type I bursts would show harder spectra
and would not be bright enough to be detected by \xmm EPIC. The
bursts identify the sources as neutron star low mass XRBs in \m31. These type I
X-ray bursts are the first detected outside the Milky Way and show that, with the
large collecting area of \xmm, X-ray bursts can be used to classify neutron star low mass XRBs in
Local Group galaxies.
 
\section{Optical novae as major class of SSS in M 31}
\begin{figure}
\centering
\includegraphics[height=8.0cm,angle=-90,clip]{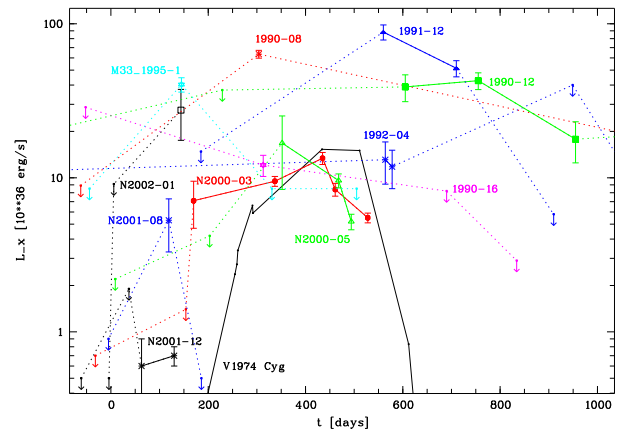}
\caption{Light curves for \m31\ and \me33\ novae that were detected within 
     1000~d after outburst. 
     Detections of individual novae are connected by solid lines, 
     and connections to upper 
     limits are marked by dashed lines 
     \citep[Fig.~3 from][]{2005A&A...442..879P}.
  \label{fig:novae}}
\end{figure}

PFF2005 searched for X-ray counterparts to optical novae 
detected in \m31\ and \me33.  
They combined an \m31\ optical nova catalogue from the WeCAPP survey with optical novae
reported in the literature and correlated them with the most recent X-ray catalogues
from \ro, \xmm\ and \chandra, and -- in addition -- searched for nova correlations 
in archival data. They report 21 X-ray counterparts for novae in \m31\ (mostly SSS). 
Their sample more than triples the number of known optical novae 
with super-soft phase. 
Most of the counterparts are covered in several observations which allows to
constrain X-ray light curves of optical novae (see Fig.~\ref{fig:novae}). 
Selected brighter sources were
classified by their \xmm\ EPIC spectra. Six counterparts are only detected in
\chandra\ HRC~I (3) or ROSAT HRI (3) observations, i.e. X-ray detectors with no
energy resolution, and therefore can not be classified as super-soft.
From the well-determined start time of the SSS state in two novae
one can estimate the hydrogen mass ejected in the outburst to 
$\sim10^{-5}M_{\odot}$ and 
$\sim10^{-6}M_{\odot}$, respectively.
The super-soft X-ray phase of at least 15\% of the novae starts within a year. 
At least one of the novae shows a SSS state lasting 6.1 years after
the optical outburst. Six of the SSSs turned on between 3 and 9 years 
after the optical discovery of the outburst and may be interpreted as 
recurrent novae. If
confirmed, the detection of a delayed SSS phase turn-on may be used as a new method 
to classify novae as recurrent. At the moment, the new method yields a
ratio of recurrent novae to classical novae of 0.3. Ongoing optical and X-ray
monitoring of the central region of \m31, where most of the novae are detected, 
should allow us to determine the length of the plateau phase of several novae
and, together with the nova temperature development, give a handle on the masses
of the white dwarfs involved.

\section{Conclusions}
The sensitivity of \xmm\ and \chandra\  observations of \m31\ combined
with the wealth of multi-wavelength data for the galaxy allows a detailed
study of the point source population. Many more interesting results can be
expected from further monitoring of \m31\  with \xmm\ and \chandra\ specifically
also in the energy band below 0.5 keV. The first light curves of the 
SSS state of optical novae proved 
that these kind of studies can be more efficiently achieved
by observing many candidates at the same time in one field in \m31\ than 
by monitoring individual novae in the Milky Way or the Magellanic
Clouds. The results of the \chandra\ and \xmm\
observations of \m31\ demonstrate the importance of 
arcsec spatial resolution, broad energy coverage, good energy resolution, 
and high collecting power -- used 
together with deep images and catalogues at other wavelengths -- also for future X-ray
source population studies in nearby galaxies.

\section*{Acknowledgments}
Based on observations obtained with 
XMM-Newton, an ESA science mission with instruments and contributions 
directly funded by ESA Member States and NASA.
The \xmm\ project is supported by the Bundesministerium f\"{u}r
Bildung und Forschung / Deutsches Zentrum f\"{u}r Luft- und Raumfahrt 
(BMBF/DLR), the Max-Planck Society and the Heidenhain-Stiftung.

\bibliographystyle{XrU2005}
\bibliography{./pietsch,/home/wnp/data1/papers/my1990,/home/wnp/data1/papers/my2000,/home/wnp/data1/papers/my2005}

\end{document}